# Discovery of the high-entropy carbide ceramic topological superconductor candidate (Ti$_{0.2}$Zr$_{0.2}$Nb$_{0.2}$Hf$_{0.2}$Ta$_{0.2}$)C

*Lingyong Zeng[1,#], Zequan Wang[2,#], Jing Song[3,#], Gaoting Lin[4], Ruixin Guo[5,6], Si-Chun Luo[7], Shu Guo[5,6], Kuan Li[1], Peifei Yu[1], Chao Zhang[1], Wei-Ming Guo[7], Jie Ma[4,8], Yusheng Hou[2,*], Huixia Luo[1,*]*

[1]School of Materials Science and Engineering, State Key Laboratory of Optoelectronic Materials and Technologies, Key Lab of Polymer Composite & Functional Materials, Guangzhou Key Laboratory of Flexible Electronic Materials and Wearable Devices

Sun Yat-Sen University, Guangzhou, 510275, China

[2]Guangdong Provincial Key Laboratory of Magnetoelectric Physics and Devices, Center for Neutron Science and Technology, School of Physics

Sun Yat-Sen University, Guangzhou, 510275, China

[3]Beijing National Laboratory for Condensed Matter Physics, Institute of Physics

Chinese Academy of Sciences, Beijing 100190, China.

[4]Key Laboratory of Artificial Structures and Quantum Control, Shenyang National Laboratory for Materials Science, School of Physics and Astronomy, Shanghai Jiao Tong University, Shanghai, 200240, China.

[5]Shenzhen Institute for Quantum Science and Engineering, Southern University of Science and Technology, Shenzhen 518055, China

[6]International Quantum Academy, Shenzhen 518048, China

[7]School of Electromechanical Engineering, Guangdong University of Technology, Guangzhou 510006, China

[8]Wuhan National High Magnetic Field Center, Huazhong University of Science and Technology, Wuhan 430074, China

[#] These authors contributed equally to this work.
*Corresponding author/authors complete details (Telephone; E-mail:) (+86)-2039386124; E-mail address: houysh@mail.sysu.edu.cn;luohx7@mail.sysu.edu.cn

**Abstract:** High-entropy ceramics (HECs) are solid solutions of inorganic compounds with one or more Wyckoff sites shared by equal or near-equal atomic ratios of multi-principal elements. Material design and property tailoring possibilities emerge from this new class of materials. Here, we report the discovery of superconductivity around 2.35 K and topological properties in the $(Ti_{0.2}Zr_{0.2}Nb_{0.2}Hf_{0.2}Ta_{0.2})C$ high-entropy carbide ceramic (HECC), which has not been observed before in any of the investigated HECC. Density functional theory calculations showed that six type-II Dirac points exist in $(Ti_{0.2}Zr_{0.2}Nb_{0.2}Hf_{0.2}Ta_{0.2})C$, which mainly contributed from the $t_{2g}$ orbitals of transition metals and the $p$ orbitals of C. Due to the stability of the structure, we also observed robust superconductivity under pressure in this HEC superconductor. This study expands the physical properties of HECs, which may become a new material platform for superconductivity research, especially for studying the coupling between superconductivity and topological physics.

## 1. Introduction

High-entropy materials (HEMs) have a highly disordered homogeneous crystalline single phase with simple lattice structures, and the metal atoms are usually composed of equimolar (or nearly equal) amounts of at least five elements.[1] The HEMs include high-entropy alloys (HEAs), high-entropy films (HEFs), high-entropy ceramics (HECs), etc.[2-4] Entropy is thought to play a key stabilizing role in these materials, providing a new perspective to develop the advanced materials.[5] Since the concept of HEAs was proposed in 2004,[6,7] HEAs have attracted widespread interest due to their potential to provide exceptional mechanical and physical functionalities, such as high strength,[8] high hardness,[9] and good corrosion resistance.[10] Furthermore, the first HEA superconductor was discovered in the Ti-Zr-Nb-Hf-Ta system in 2014,[11] revealing a new aspect of the capabilities of HEAs. Since then, there have been numerous studies performed on HEA superconductors. In $(HfZrTi)_x(TaNb)_{1-x}$ ($0.2 \leq x \leq 0.84$) systems, the superconducting transition temperatures ($T_c$s) have a strong correlation with the valence electron count (VEC).[12] In particular, the $(HfZrTi)_{0.33}(TaNb)_{0.67}$ sample exhibits a robust SC against a high physical pressure of up to 190.6 GPa.[13] Besides, the superconducting properties of $(HfZrTi)_x(TaNb)_{1-x}$ HEFs are also reported.[14] The currently known HEA superconductors are also found in hcp, CsCl, and $a$-Mn crystal structures.[15-18] Recently, the concept of high-entropy has been applied to a copper-based high-temperature superconductor (REBa$_2$Cu$_3$O$_{7-\delta}$, RE = Y, La, Nd, Sm, Eu, and Gd),[19] and NaCl-type telluride compound (AgInSnPbBiTe$_5$).[20] However, there are no reports on the superconductivity (SC) in high-entropy carbide ceramics (HECCs).

The binary transition-metal monocarbides (TMMCs) have been widely studied due to their exotic superconducting and topological properties. The topological surface states were observed in WC, and the non-trivial topology of its newly identified semimetal was also demonstrated.[21] The first-principles calculation indicates that *s*-wave Bardeen-Cooper-Schrieffer (BCS) SC and non-trivial band topology coexist in cubic *a*-MoC and hexagonal *γ*-MoC.[22] Besides, NbC was found to be a type-II 11.5 K superconductor with topological non-trivial surface states, supporting NbC as a potential topological superconductor.[23] Moreover, the type-II Dirac semimetal states were proposed to exist in the band structure of TaC, which is a relatively high $T_c$ ~ 10.6 K superconductor.[24,25] However, TiC, ZrC, and HfC, which have 4 VEC per atom, are non-superconducting.[26]

HECs have entered the fray and garnered increasing interest from 2015 onward.[27] The HECs are defined as the solid solution of five or more cations or anions sublattices with a high configuration entropy.[3,27-30] Combining the advantages of high-entropy with the novel physical properties in TMMCs, high-entropy carbides ceramics (HECCs) may be a new platform for studying the coupling between SC and topological physics. Furthermore, the different compositions of transition metal elements in HECs can present various VECs, which provides a good carrier for studying the relationship between VEC and $T_c$ and is expected to prepare high-$T_c$ superconductors in HECCs based on the Matthias rule.[31]

Here we reported the SC in the HECC (Ti$_{0.2}$Zr$_{0.2}$Nb$_{0.2}$Hf$_{0.2}$Ta$_{0.2}$)C for the first time. The superconducting and topological properties of (Ti$_{0.2}$Zr$_{0.2}$Nb$_{0.2}$Hf$_{0.2}$Ta$_{0.2}$)C were investigated experimentally and computationally. This material displays bulk type-II SC with $T_c$ of about 2.35 K, and a series of superconducting parameters have been determined. By performing high-pressure resistance measurements, we found that the SC in this HEC sample is robust. Density functional theory (DFT) calculations show that (Ti$_{0.2}$Zr$_{0.2}$Nb$_{0.2}$Hf$_{0.2}$Ta$_{0.2}$)C has six type-II Dirac points mainly contributed from the $t_{2g}$ orbitals of transition metals, and the *p* orbitals of C. The coexistence of the SC and Dirac points in (Ti$_{0.2}$Zr$_{0.2}$Nb$_{0.2}$Hf$_{0.2}$Ta$_{0.2}$)C suggests that it is a candidate for topological superconductors. Our work provides a new materials platform for studying the coupling between superconductivity and topological physics.

## 2. Results and Discussions

Figure 1a shows the detailed refinement results of $(Ti_{0.2}Zr_{0.2}Nb_{0.2}Hf_{0.2}Ta_{0.2})C$ HEC. The powder X-ray diffraction (PXRD) pattern can all be indexed with a face-centered cubic (fcc) rock-salt structure with space group $Fm\bar{3}m$, which is consistent with the structure of the five TMMCs binary.[32] No impurity phases could be detected, indicating a good sample quality. The lattice parameters from the Rietveld fitted PXRD profile are a = b = c = 4.51711(9) Å, close to the previously reported value.[28] Figure 1b shows the rock-salt crystal structure of this HEC sample. The five transition metal elements are solid solutions in a cation position, while the carbon element occupies the anion position. SEM-EDS results from Figure 1c and Figure S1 reveal the actual element ratio is close to the design value.

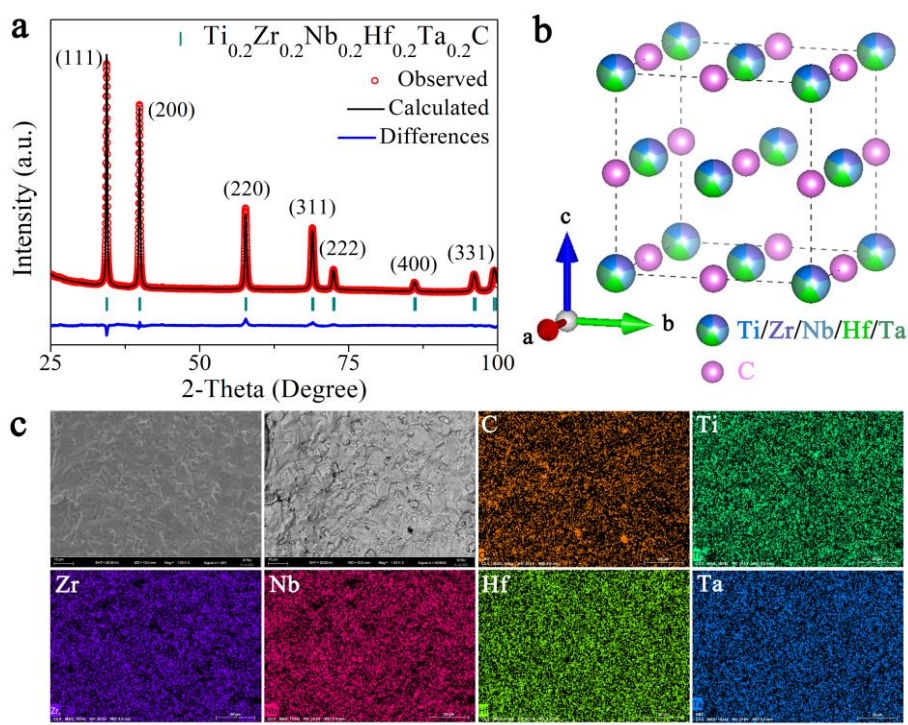

**Figure 1.** Structural and microstructural characterization of $(Ti_{0.2}Zr_{0.2}Nb_{0.2}Hf_{0.2}Ta_{0.2})C$ HEC. a) The Rietveld refinement pattern of HEC $(Ti_{0.2}Zr_{0.2}Nb_{0.2}Hf_{0.2}Ta_{0.2})C$. b) The crystal structure of HEC $(Ti_{0.2}Zr_{0.2}Nb_{0.2}Hf_{0.2}Ta_{0.2})C$. c) SEM and BSE image of the fracture surface and the corresponding EDS mapping for $(Ti_{0.2}Zr_{0.2}Hf_{0.2}Nb_{0.2}Ta_{0.2})C$.

To characterize the new HEC superconductor, the field-dependent volume magnetization is collected at 1.8 K and linear fits ($M_{fit}$ = e + fH) in the low field region. The demagnetization N was determined with the formula -f = $1/(4\pi(1-N))$. We further characterized through zero-field-cooled (ZFC) magnetic susceptibility measurements under a 2 mT magnetic field. Figure 2a shows the magnetic susceptibility data corrected with N values. As indicated by the red arrow in Figure 2a, a clear diamagnetic signal appears below the $T_c$ of 2.35 K. The ZFC measurement was corrected for N = 0.56. The resulting diamagnetic signal is very close to the ideal value of

-1. The inset in Fig. 2a shows field-dependent magnetization measurements on the HEC superconductor (Ti$_{0.2}$Zr$_{0.2}$Nb$_{0.2}$Hf$_{0.2}$Ta$_{0.2}$)C at different temperatures below $T_c$. Figure S2 shows the M-M$_{fit}$ curves. The field at which the magnetization begins to deviate from a linear response is the uncorrected lower critical field, $\mu_0 H_{c1}^*$, for that temperature. Figure 2b shows the $\mu_0 H_{c1}^*(T)$ points and fitted to the formula: $\mu_0 H_{c1}^*(T) = \mu_0 H_{c1}^*(0)(1-(T/T_c)^2)$. The $\mu_0 H_{c1}^*(0)$ was calculated as 11.5 mT for (Ti$_{0.2}$Zr$_{0.2}$Nb$_{0.2}$Hf$_{0.2}$Ta$_{0.2}$)C. After correcting for N = 0.56, $\mu_0 H_{c1}(0)$ was calculated to be 26.1 mT.

Figure S3 shows the $\rho(T)/\rho(300K)$ data for three (Ti$_{0.2}$Zr$_{0.2}$Nb$_{0.2}$Hf$_{0.2}$Ta$_{0.2}$)C sample pieces. There is a sudden drop in resistivity near $T_c$. The zero resistance temperature is 2.35 K, which agrees well with the $T_c \sim$ 2.35 K observed from magnetic susceptibility. The resistivity data at various fixed magnetic fields for this HEC superconductor are presented in Figure 2c. Figure 2d shows the H$_{c2}$-T phase diagram of the new HEC superconductor. The estimated $T_c$ values under different magnetic fields can be fitted to a line, giving the slope $d\mu_0 H_{c2}/dT$ = -0.224. The upper critical field $\mu_0 H_{c2}(0)$ was determined through the Werthamer-Helfand-Hohenberg (WHH) formula: $\mu_0 H_{c2}(0) = -0.693 T_c (\frac{d\mu_0 H_{c2}}{dT})|_{T=T_c}$. The dirty limit $\mu_0 H_{c2}(0)$ was calculated as 0.39 T using $T_c$ = 2.53 K (50% $\rho_N$). We also calculated the $\mu_0 H_{c2}(0)$ with the Ginzburg-Landau (GL) equation: $\mu_0 H_{c2}(T) = \mu_0 H_{c2}(0) * \frac{1-(T/T_c)^2}{1+(T/T_c)^2}$. Fitting with the data can yield $\mu_0 H_{c2}(0)^{GL}$ = 0.51 T. Furthermore, the H$_{c2}$(0)/$T_c$ ratio is about 0.217 T/K for (Ti$_{0.2}$Zr$_{0.2}$Nb$_{0.2}$Hf$_{0.2}$Ta$_{0.2}$)C HEC, much larger than 0.168 T/K and 0.063 T/K for NbC and TaC compounds. Thus, the pairing interaction is more robust in this HEC superconductor. Nonetheless, it is noted that the H$_{c2}$(0)/$T_c$ value of this HEC superconductor is far below 1.86 expected for the Pauli paramagnetic limit, implying that the orbital effect limits SC.[33]

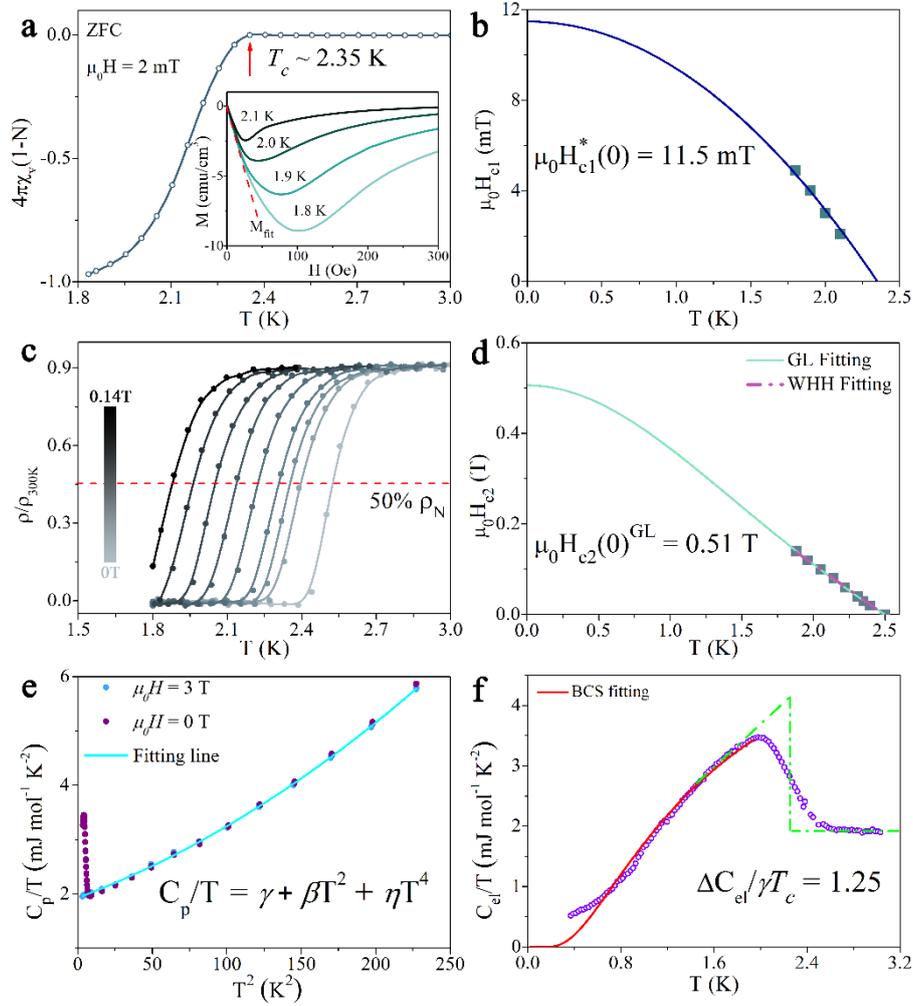

**Figure 2.** The superconductivity measurements of $(Ti_{0.2}Zr_{0.2}Nb_{0.2}Hf_{0.2}Ta_{0.2})C$ HEC. a) Magnetization behavior of this HEC superconductor measured under a 2 mT magnetic field. The inset shows field-dependent volume magnetization measured at various temperatures below $T_c$. b) Lower critical field $\mu_0 H_{c1}^*(0)$ versus temperature. c) The resistive transition under different magnetic fields. d) The $H_{c2}$-T phase diagram of this HEC superconductor. e) Temperature dependence of specific heat divided by $T^2$ for $(Ti_{0.2}Zr_{0.2}Nb_{0.2}Hf_{0.2}Ta_{0.2})C$ in 0 and 3 T field. f) The electronic specific heat of this HEC superconductor.

After observing the Meissner effect and the zero resistivity, we measured the heat capacity under 0 and 3 T magnetic fields to fully confirm the bulk nature of the SC in $(Ti_{0.2}Zr_{0.2}Nb_{0.2}Hf_{0.2}Ta_{0.2})C$ superconductor. As shown in Figure 2e, an apparent anomaly in the 0 T heat capacity corresponds to the superconducting state's appearance around 2.35 K, consistent with resistivity and magnetization measurements. Under the magnetic field of 3 T, the superconducting heat capacity jump has disappeared. The non-superconducting state heat capacity can be fitted according to the relation: $C_p/T = \gamma + \beta T^2 + \eta T^4$, where two-term, $\beta T^2 +$ $\eta T^4$, are used to express the phonon contribution, and $\gamma$ represents the normal state electronic specific heat coefficient. Such fitting has been applied to TMMC superconductors,[25] Kagome lattice superconductors,[34] and some unconventional superconductors[35] as well. This relation matches the experimental data well and yields $\gamma$ = 1.85 mJ mol$^{-1}$ K$^{-2}$ and $\beta$ = 0.0102 mJ mol$^{-1}$ K$^{-4}$. Figure 2f presents the normalized electronic specific data for $(Ti_{0.2}Zr_{0.2}Nb_{0.2}Hf_{0.2}Ta_{0.2})C$ HEC. The normalized heat capacity jump ($\Delta C/\gamma T_c$) is calculated as 1.25 based on the equal area method.

In addition, we calculated the Debye temperature based on the formula $\Theta_D = (12\pi^4 nR/5\beta)^{1/3}$, where R is the gas constant and $n$ represents the number of atoms per formula unit ($n$ = 2 for $(Ti_{0.2}Zr_{0.2}Nb_{0.2}Hf_{0.2}Ta_{0.2})C$), it gives $\Theta_D$ = 724 K. The electron-phonon coupling constant ($\lambda_{ep}$) can then be estimated through the inverted McMillan formula, $\lambda_{ep} = \dfrac{1.04 + \mu^* \ln\left(\dfrac{\Theta_D}{1.45 T_c}\right)}{(1 - 0.62\mu^*) \ln\left(\dfrac{\Theta_D}{1.45 T_c}\right) - 1.04}$,[36] where $\mu^*$ is the Coulomb pseudopotential parameter and is typically given a value of 0.13.[37,38] The $\lambda_{ep}$ is calculated as 0.45, indicating weakly coupled SC. Finally, the electron DOS at Fermi level ($N(E_F)$) can be estimated by using the relation $N(E_F) = \dfrac{3}{\pi^2 k_B^2 (1 + \lambda_{ep})}\gamma$. The calculated $N(E_F)$ is around 0.54 states/eV f.u, lower than 0.8 for TaC and NbC.[25] A summary of observations and estimates of superconducting parameters is provided in Table S1.

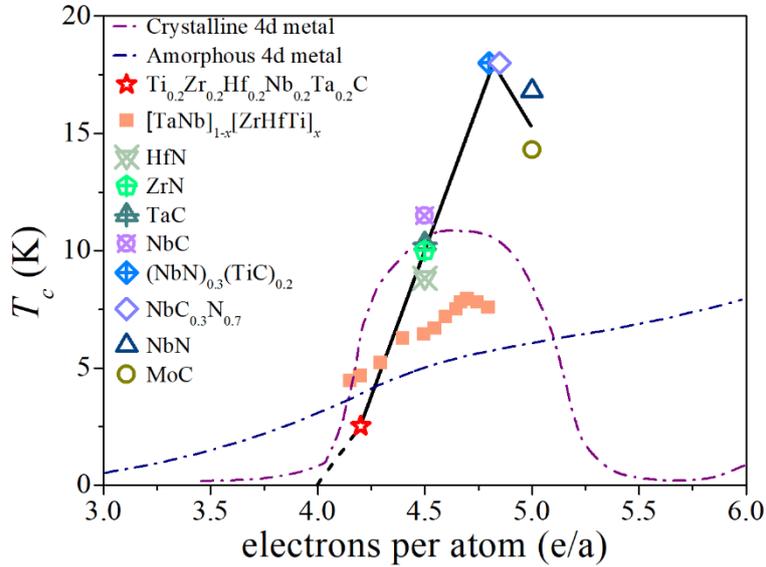

**Figure 3.** Temperature-e/a phase diagram. VEC per atom dependency of the $T_c$ for transition metal carbides and nitrides, $[TaNb]_{1-x}[ZrHfTi]_x$ systems and $(Ti_{0.2}Zr_{0.2}Nb_{0.2}Hf_{0.2}Ta_{0.2})C$.[12,22-24,26,31,39-41]

Figure 3 displays the $T_c$ of the transition metal carbides and nitrides and $[TaNb]_{1-x}[ZrHfTi]_x$ systems as a function of the electron/atom (e/a) ratio. The trend lines of the $T_c$ for the transition metals and their alloys in crystalline form and amorphous vapor-deposited film are also described.[31,39] It is often referred to as the Matthias rule, which shows that the $T_c$ maximum occurs between 4.7 and 6.5 e/a for transition metals. Compared with the structures between the TMMCs and transition metal nitrides (TMMNs) and (Ti$_{0.2}$Zr$_{0.2}$Nb$_{0.2}$Hf$_{0.2}$Ta$_{0.2}$)C HEC, they have the same NaCl-type crystal structure. The IVB-IVA TMMC compounds TiC, ZrC, and HfC, which have 4 e/a, are non-superconducting.[26] When the e/a increase to 4.2 of (Ti$_{0.2}$Zr$_{0.2}$Nb$_{0.2}$Hf$_{0.2}$Ta$_{0.2}$)C HEC, the $T_c$ is 2.35 K. The VB-IVA and VB-VA compounds NbC, TaC, ZrN, and HfN have 4.5 e/a and high $T_c$ values.[24,40] The highest $T_c$ found up to now for transition metal carbides, and nitrides are about 18 K for pseudo-binary systems (NbN)$_{0.8}$(TiC)$_{0.2}$ and NbC$_{0.3}$N$_{0.7}$,[26] which have 4.8 and 4.85 e/a, respectively. According to Matthias' rule[31] and the study of SC in interstitial compounds with NaCl-type structure,[26] when the electrons per atom are higher than 4.8, the $T_c$ decreases. Thus, MoC and NbN with 5 e/a have the $T_c$ of 16.8 K and 14.3 K, respectively.[22,41] However, producing pseudo-binary systems with about 4.8 electrons per atom is difficult since they show lattice instabilities and undergo a structural phase transition.[42] While in HECCs, the high configurational entropy of chemically complex solution phases stabilizes them into a single crystalline and exhibits better stability, especially in HECCs with the high entropy-forming ability (EFA) and minor transition-metal lattice size difference.[5,29] And due to the infinite possibilities in composition design (for increasing the number of e/a) and property tuning paradigms, HECCs may be the next good candidate to achieve high $T_c$ in transition metal carbides.

We performed high-pressure, high-pressure resistance measurements to investigate the SC's robustness in (Ti$_{0.2}$Zr$_{0.2}$Nb$_{0.2}$Hf$_{0.2}$Ta$_{0.2}$)C HEC. Figure 4a shows the R(T) data for pressures between 0.6 and 80.3 GPa. The superconducting transitions of this HEC superconductor subjected to different pressures are sharp. Focusing on the low-temperature region R(T) curves, it is seen that the $T_c$ shifts to lower temperature monotonically as pressure increases (see Figure 4b). Unfortunately, we could not see a full SC transition with zero resistance because it is out of the low-T limit of our cryostat. The pressure-dependent $T_c$'s are summarized in the pressure-temperature phase diagram in Figure 4c. $T_c$ exhibits a slight decrease from its ambient-pressure value of ~ 2.67 K to 2.15 K at ~ 50 GPa. When the pressure increases, SC transition shifts below 2.15 K, cut off by the low T limit until ~ 80 GPa. We see a robust SC subject to pressure, with a slight $T_c$ variation of 0.5 K within 50 GPa, similar to HEA, unlike typical SC (Pb) (~5 K)[43] or unconventional SC (cuprate, iron-based SC, etc.) (~20 K).[44, 45] High pressure (HP) XRD measurements of (Ti$_{0.2}$Zr$_{0.2}$Nb$_{0.2}$Hf$_{0.2}$Ta$_{0.2}$)C HECC showed that no structural phase transition is observed to at least 18 GPa, as shown in Figure S4. The robustness of SC in this HEC superconductor could be related to its outstanding structural stability under pressure.

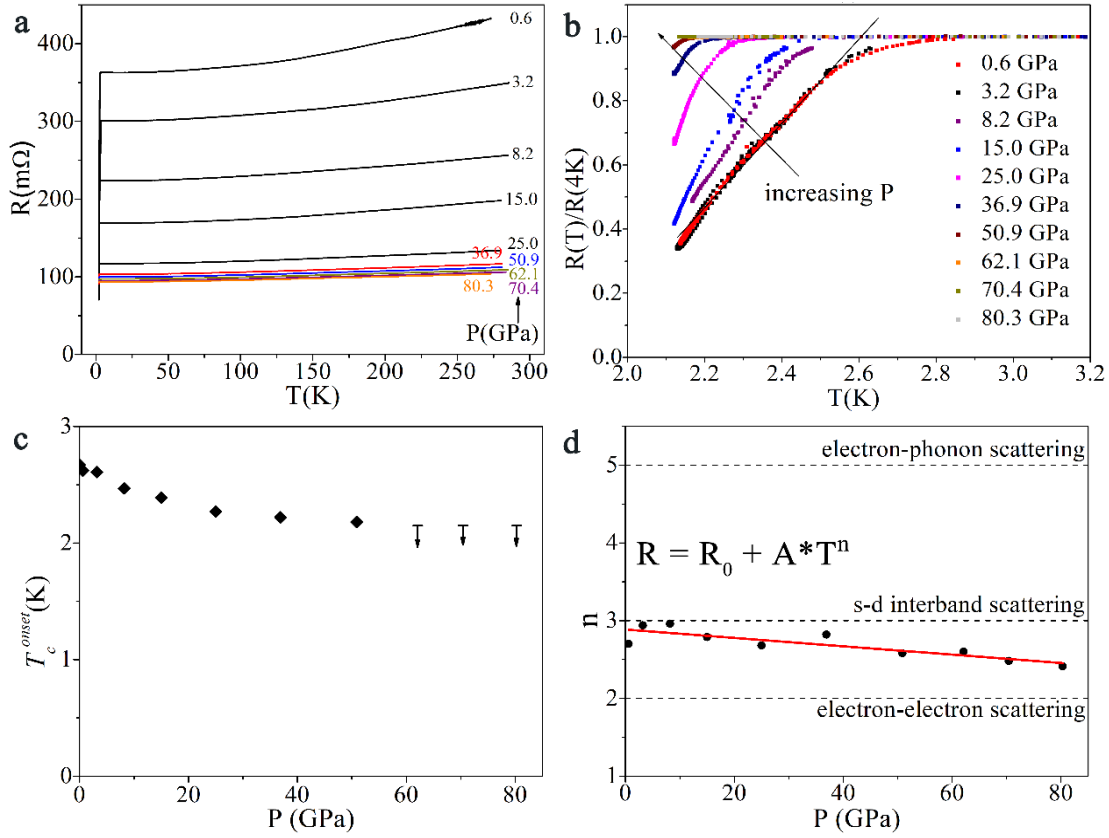

**Figure 4.** Transport properties of the $(Ti_{0.2}Zr_{0.2}Nb_{0.2}Hf_{0.2}Ta_{0.2})C$ HEC. a) Resistance of $(Ti_{0.2}Zr_{0.2}Nb_{0.2}Hf_{0.2}Ta_{0.2})C$ HEC versus temperature from 2.15 to 280 K for pressures to 80.3 GPa. b) Resistance normalized against resistance at 4 K for temperatures 2.15 to 3.2 K. Here $T_c^{onset}$ is defined as the temperature where the straight line hits the residual resistance. c) $T_c^{onset}$ of $(Ti_{0.2}Zr_{0.2}Nb_{0.2}Hf_{0.2}Ta_{0.2})C$ HEC versus pressure. Vertical arrows indicate SC transitions below the cryostat's temperature limit (2.15 K). d) Pressure dependence of the power law exponent. The resistance below 50 K is fitted with the $R = AT^n + R_0$, where A represents the power law coefficient, $R_0$ represents the residual resistance, and n is the exponent of the power law.

The T-dependent resistance of $(Ti_{0.2}Zr_{0.2}Nb_{0.2}Hf_{0.2}Ta_{0.2})C$ HEC shows interesting scaling behaviors (see Figure 4d). When fitted with the following equation $R = A*T^n + R_0$ below 50 K, A represents the power law coefficient, $R_0$ is the residual resistance, and n is the exponent. It gives a value of the power law exponent n close to 3, which is different from the value expected for a system dominated by either electron-electron scattering (n = 2) or electron-phonon scattering (n = 5). Considering the Debye temperature (~ 724 K) obtained from the specific heat measurements is far above 50 K, the n value near 3 indicates that the s–d interband scattering dominates the electron transport with contributions from electron correlation effects, as observed in 1T-TiSe$_2$[46] or Ta$_4$Pd$_3$Te$_{16}$[47]. Upon further applying pressure, n starts to decrease towards 2, in agreement with the expectation that the *s* band shifts to higher energy,[48] promoting the *s-d* electron transfer, increasing the number of d electrons, and strengthening the electron correlation effects.

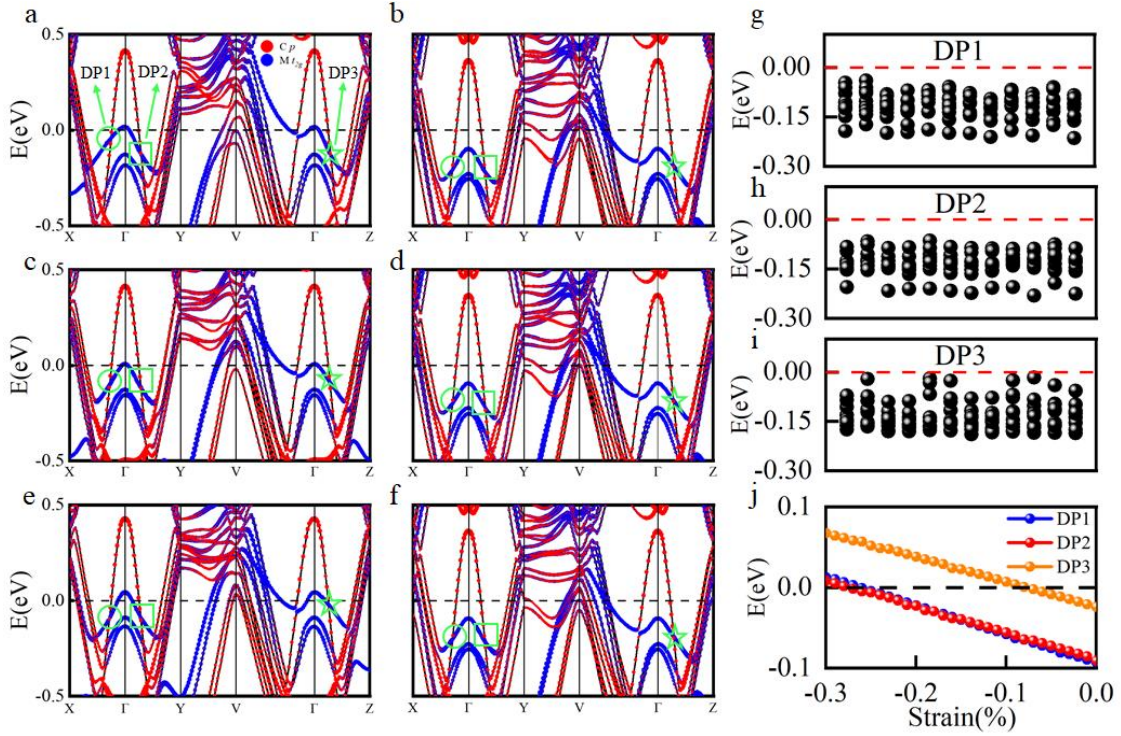

**Figure 5.** DFT calculated the electronic properties of (Ti$_{0.2}$Zr$_{0.2}$Nb$_{0.2}$Hf$_{0.2}$Ta$_{0.2}$)C. a)-f) Orbital-resolved band structures of six representative SQSs when SOC is considered in DFT calculations. The green arrows indicate the type-II DPs. The green circles, squares, and stars represent DP1, DP2, and DP3 respectively. The Fermi level is set to zero in a)-f). The energies of the type-II DPs when they are located along X-Γ g), Γ-Y h), and Γ-Z i). j) The dependence of the energies of the type-II DPs on the compressive strain in (Ti$_{0.2}$Zr$_{0.2}$Nb$_{0.2}$Hf$_{0.2}$Ta$_{0.2}$)C.

To gain a deeper insight into the topological properties of (Ti$_{0.2}$Zr$_{0.2}$Nb$_{0.2}$Hf$_{0.2}$Ta$_{0.2}$)C, we employed DFT calculations to study its electronic properties. Figure 5a-f show orbital-resolved band structures of six representative special quasi-random structures (SQSs) of (Ti$_{0.2}$Zr$_{0.2}$Nb$_{0.2}$Hf$_{0.2}$Ta$_{0.2}$)C when spin-orbital coupling (SOC) is considered. The crystal structures of these six representative SQSs and their first Brillouin zones are given in Figure S5. First, we see Dirac points (DPs) located along X-Γ, Γ-Y, and Γ-Z. Because these DPs tilt and have the same sign of velocities along X-Γ, Y-Γ, and Z-Γ, they form the type-II DPs. Additionally, our DFT calculations show that (Ti$_{0.2}$Zr$_{0.2}$Nb$_{0.2}$Hf$_{0.2}$Ta$_{0.2}$)C has six type-II DPs (see Figure S6). By projecting the wave functions of bands to transition metals M (M=Ti, Zr,

Nb, Hf, Ta) and C, it shows that the type-II DPs are mainly contributed from two nondegenerate bands, i.e., one is from the $t_{2g}$ orbitals of transition metals M while the other is from the $p$ orbitals of C. Such a feature indicates that the type-II DPs in the HEC $(Ti_{0.2}Zr_{0.2}Nb_{0.2}Hf_{0.2}Ta_{0.2})C$ are closely related to the band inversion at Γ point, which is similar to the superconductor TaC.[49] It is worth noting that $(Ti_{0.2}Zr_{0.2}Nb_{0.2}Hf_{0.2}Ta_{0.2})C$ exhibits three nodal loops when SOC is not considered in DFT calculations (Figure S7-S9). Combining the experimentally observed superconductivity and topologically non-trivial DPs revealed by DFT calculations, it is plausible that $(Ti_{0.2}Zr_{0.2}Nb_{0.2}Hf_{0.2}Ta_{0.2})C$ is a possible candidate for topological superconductors.

To study the influence of the disorder from the transition metal M on the electronic properties of the HEC $(Ti_{0.2}Zr_{0.2}Nb_{0.2}Hf_{0.2}Ta_{0.2})C$, we investigate its band structures with different structural configurations. Here, we take the positions of the same transition metal elements as an individual region and then randomly place the five various transition metals M (M = Ti, Zr, Hf, Nb, Ta) on the five areas. Finally, all possible permutations between transition metals are considered to simulate the disorder of this HEC superconductor. Overall, the type-II DPs exhibit all considered structural configurations. The difference among them is the energies for the type-II DPs concerning the Fermi surface. Figures 5g, h, and i show these energies when the type-II DPs are along X-Γ, Γ-Y, and Γ-Z, respectively. It is clearly shown that the energies of the type-II DP away from the Fermi Level fluctuate between -0.231 and -0.017 eV as the disorder of the transition metals change. It is worth noting that the energies of the type-II DPs from the Fermi level are -1.30 and -1.0 eV in TaC and NbC,[23, 25, 49] respectively, which are much lower than those of $(Ti_{0.2}Zr_{0.2}Nb_{0.2}Hf_{0.2}Ta_{0.2})C$. Moreover, we find that the energies of the type-II DPs vary linearly as the compressive strain increase (Figure 5j). Interestingly, the type-II DPs along Γ-Z, Γ-Y, and X-Γ will precisely located at the Fermi level when the compressive strain is -0.08 %, -0.26 %, and -0.28 %, respectively (see Figure S10). Hence, alloying transition metals and applying compressive pressure should be helpful to realize topological superconductivity in the HEC $(Ti_{0.2}Zr_{0.2}Nb_{0.2}Hf_{0.2}Ta_{0.2})C$.

Unusual features in the specific heat data of a superconductor may indirectly hint towards unconventional and potentially topological superconductivity.[50] As we know, the behavior of specific heat below $T_c$ can be qualitatively different for conventional and unconventional superconductors. Below $T_c$, the specific heat of a superconductor can be described by the following formula: $C(T) = C_n(T) - \propto T^3 \exp(-\frac{\Delta(0)}{k_B T})$, where $C_n(T)$ is the specific heat in the normal state, α is a constant, $\Delta(0)$ is the superconducting energy gap at zero temperature, and $k_B$ is the Boltzmann constant. The exponential term in this formula reflects the suppression of low-energy excitations due to the formation of Cooper pairs. As the temperature approaches zero, the specific heat of the superconductor decreases exponentially, eventually becoming negligible. In unconventional superconductors, the specific heat behavior below $T_c$ can be deviating from the above behavior, reflecting the presence of competing orders or a non-trivial gap structure. Fig. 2f shows the normalized electronic specific data for $(Ti_{0.2}Zr_{0.2}Nb_{0.2}Hf_{0.2}Ta_{0.2})C$ HECC and the BCS theory is applied for fitting the data. The BCS theory does not fit data points very well. $C_{el}/T$ decreases much more slowly than the BCS behavior; in particular, $C_{el}/T$ keeps showing a sizable temperature dependence even at our lowest temperature of 0.36 K (T/$T_c$ = 0.15), whereas $C_{el}/T$ should already become negligible at such a low temperature in the BCS case. Such a peculiar behavior in $C_{el}/T$ suggests the existence of nodes in the superconducting gap.[51] The $C_{el}/T$ data, particularly the strong T dependence near 0 K, points to the realization of unconventional superconductivity in $(Ti_{0.2}Zr_{0.2}Nb_{0.2}Hf_{0.2}Ta_{0.2})C$ HECC. Further combined with the results of the DFT calculated, we have reason to believe that $(Ti_{0.2}Zr_{0.2}Nb_{0.2}Hf_{0.2}Ta_{0.2})C$ HECC may be a candidate for topological superconductors.

## 3. Conclusion

In summary, we discovered the SC and topological properties in HECC $(Ti_{0.2}Zr_{0.2}Nb_{0.2}Hf_{0.2}Ta_{0.2})C$, which has not been observed before in any of the investigated HECC. This material shows type-II bulk SC with $T_c$ ~ 2.35 K, $\mu_0H_{c1}(0)$ ~ 26.1 mT, and $\mu_0H_{c2}(0)$ ~ 0.51 T. And due to the infinite possibilities in composition design (for increasing the number of e/a) and property tuning paradigms, HECCs may be the next good candidate to achieve high $T_c$ in transition metal carbides. We also observed the robustness of superconductivity under pressure in the $(Ti_{0.2}Zr_{0.2}Nb_{0.2}Hf_{0.2}Ta_{0.2})C$ HEC. DFT calculations showed that six type-II Dirac points exist in this HEC material, suggesting it is a potential topological superconductor. Furthermore, our work hints that it is possible to discover more superconductors, even topological superconductors, in HECs.

## 4. Experimental Section

*$(Ti_{0.2}Zr_{0.2}Nb_{0.2}Hf_{0.2}Ta_{0.2})C$ preparation*: The $(Ti_{0.2}Zr_{0.2}Nb_{0.2}Hf_{0.2}Ta_{0.2})C$ HEC were synthesized by spark plasma sintering (SPS). The synthesis procedures of $(Ti_{0.2}Zr_{0.2}Hf_{0.2}Nb_{0.2}Ta_{0.2})C$ HECs have been reported in detail elsewhere.[30, 52]

*Structural and microstructural characterization*: The structural parameters and phase purity were characterized by PXRD using a conventional X-ray obtained over the 2$\theta$ range of 10 - 100º with a step width of 0.01º. The lattice parameters and the occupancy of the atoms were refined using the Rietveld method by adopting the NaCl-type structural model and the Fullprof suite package. The elemental composition for the final product was determined by scanning electron microscope combined with energy-dispersive X-ray spectroscopy (SEM-EDS).

*Physical properties measurements*: Electrical transport, magnetization, and specific heat measurements were measured in a Quantum Design PPMS-14T.

*High-pressure measurements*: High pressures were generated between two opposed diamond anvils with 0.3 mm culets in a diamond anvil cell made of CuBe alloy. High-pressure four-point dc electrical resistivity measurements used a thin square-shaped HEC sample placed atop four thin gold leads with a cBN-epoxy mixture to insulate the T301 stainless steel gasket and with NaCl as the pressure medium. The sample loaded a little ruby sphere (~10 μm) as a pressure manometer. It can be referred to for further details of high-pressure resistivity techniques elsewhere.[53, 54]

*First-principle theoretical calculation*: Our density functional theory (DFT) calculations are performed using the Vienna *ab initio* simulation package (VASP). The generalized gradient approximation with the Perdew-Burke-Ernzerhof exchange-correlation function is used in our DFT calculations.[55] We treat Ti$3d4s$, Zr$4d5s$, Nb$4p5s4d$, Hf$5d6s$, Ta$5d6s$, and C$2s2p$ as valance electrons and employ the projector-augmented-wave pseudopotentials to describe core-valence interactions. The plane wave cutoff energy is set to be 400 eV, and a k-mesh grid of 6 × 6 × 6 centered at the Γ-point is sampled in the Brillouin zone. The high-symmetry K-path for band structures is displayed in Figure S7-a. The total energy criterion at the self-consistent step and force convergences criteria are 1×10$^{-6}$ eV and 0.01 eV/Å, respectively. The experimentally measured lattice constant of the HEC (Ti$_{0.2}$Zr$_{0.2}$Hf$_{0.2}$Nb$_{0.2}$Ta$_{0.2}$)C is adopted in our DFT calculations. Spin-orbit coupling (SOC) is included in calculating band structures. To describe the chemical disorder in the HEC (Ti$_{0.2}$Zr$_{0.2}$Hf$_{0.2}$Nb$_{0.2}$Ta$_{0.2}$)C, we utilize the special quasi-random structure (SQS) with 64 atoms,[56,57] which the Monte Carlo Special Quasirandom Structure obtains in the Alloy Theoretic Automated Toolkit.

**Supporting Information**

Supporting Information is available from the Wiley Online Library or the author.

**Acknowledgments**

The authors acknowledge productive conversations with Robert J. Cava at Princeton University. L. Y. Zeng, Z. Q. Wang, and J. Song contributed equally to this work. This work is supported

by the National Natural Science Foundation of China (12274471, 11922415, 12104518, 22205091), Guangdong Basic and Applied Basic Research Foundation (2022A1515011168, 2019A1515011718, 2022A1515012643), Guangzhou Basic and Applied Basic Research Foundation (202201011118), the Key Research & Development Program of Guangdong Province, China (2019B110209003), the Pearl River Scholarship Program of Guangdong Province Universities and Colleges (20191001). J.S. is thankful for the funding from the Institute of Physics Chinese Academy of Sciences Sta rtup Funds (BR202118) and the National Natural Science Foundation of China (12204514). J. Ma is thankful for the funding of the interdisciplinary program at Wuhan National High Magnetic Field Center (Grant No. WHMFC 202122), Huazhong University of Science and Technology. High-pressure synchrotron XRD work was performed at beamline 13-BM-C of the Advanced Photon Source (APS), Argonne National Laboratory, in the USA.

Received: ((will be filled in by the editorial staff))
Revised: ((will be filled in by the editorial staff))
Published online: ((will be filled in by the editorial staff))

# Supporting Information

**Discovery of the high-entropy carbide ceramic topological superconductor candidate (Ti$_{0.2}$Zr$_{0.2}$Nb$_{0.2}$Hf$_{0.2}$Ta$_{0.2}$)C**

*Lingyong Zeng$^{1,\#}$, Zequan Wang$^{2,\#}$, Jing Song$^{3,\#}$, Gaoting Lin$^{4}$, Ruixin Guo$^{5,6}$, Si-Chun Luo$^{7}$, Shu Guo$^{5,6}$, Kuan Li$^{1}$, Peifei Yu$^{1}$, Chao Zhang$^{1}$, Wei-Ming Guo$^{7}$, Jie Ma$^{4,8}$, Yusheng Hou$^{2,*}$, Huixia Luo$^{1,*}$*

[1]School of Materials Science and Engineering, State Key Laboratory of Optoelectronic Materials and Technologies, Key Lab of Polymer Composite & Functional Materials, Guangzhou Key Laboratory of Flexible Electronic Materials and Wearable Devices
Sun Yat-Sen University, Guangzhou, 510275, China
E-mail: luohx7@mail.sysu.edu.cn

[2]Guangdong Provincial Key Laboratory of Magnetoelectric Physics and Devices, Center for Neutron Science and Technology, School of Physics
Sun Yat-Sen University, Guangzhou, 510275, China
E-mail: houysh@mail.sysu.edu.cn

[3]Beijing National Laboratory for Condensed Matter Physics, Institute of Physics
Chinese Academy of Sciences, Beijing 100190, China.

[4]Key Laboratory of Artificial Structures and Quantum Control, Shenyang National Laboratory for Materials Science, School of Physics and Astronomy, Shanghai Jiao Tong University, Shanghai, 200240, China.

[5]Shenzhen Institute for Quantum Science and Engineering, Southern University of Science and Technology, Shenzhen 518055, China

[6]International Quantum Academy, Shenzhen 518048, China

[7]School of Electromechanical Engineering, Guangdong University of Technology, Guangzhou 510006, China

[8]Wuhan National High Magnetic Field Center, Huazhong University of Science and Technology, Wuhan 430074, China

[#] These authors contributed equally to this work.
*Corresponding author/authors complete details (Telephone; E-mail:) (+86)-2039386124; E-mail address: houysh@mail.sysu.edu.cn;luohx7@mail.sysu.edu.cn

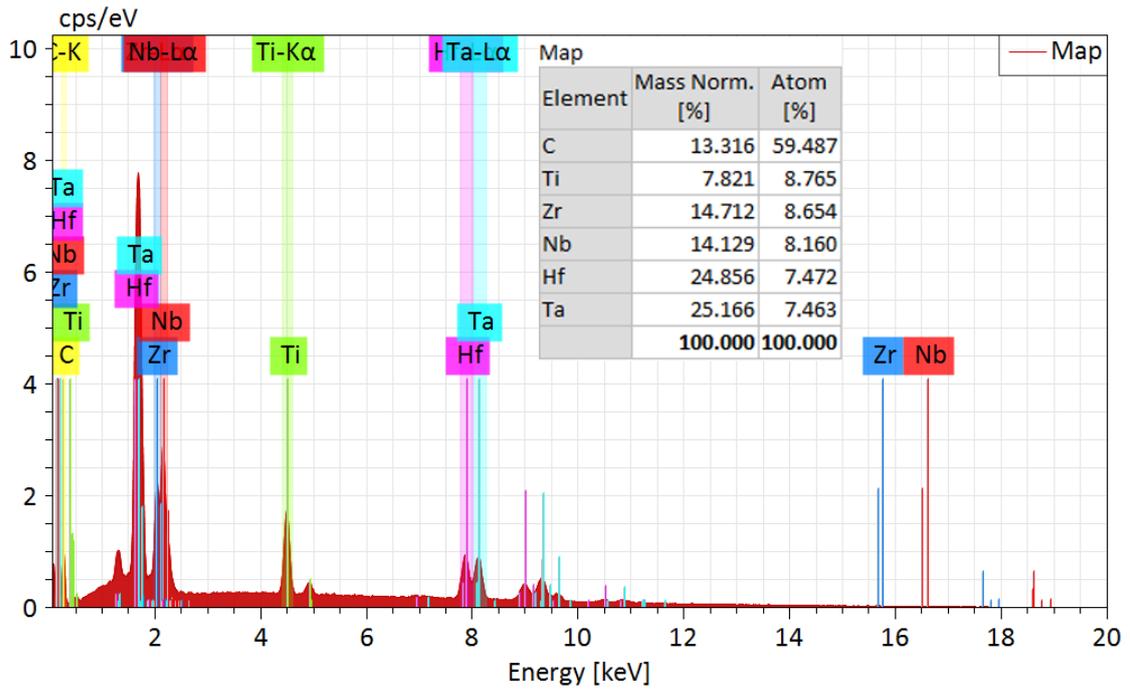

**Figure S1.** The EDS spectrum of $(Ti_{0.2}Zr_{0.2}Hf_{0.2}Nb_{0.2}Ta_{0.2})C$.

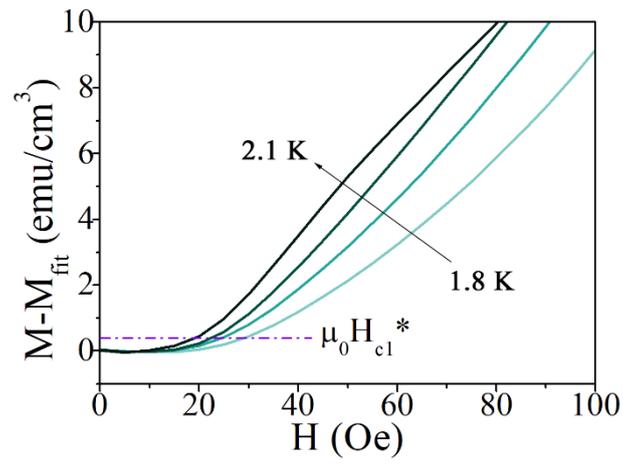

**Figure S2.** The $M-M_{fit}$ versus field curves of $(Ti_{0.2}Zr_{0.2}Hf_{0.2}Nb_{0.2}Ta_{0.2})C$.

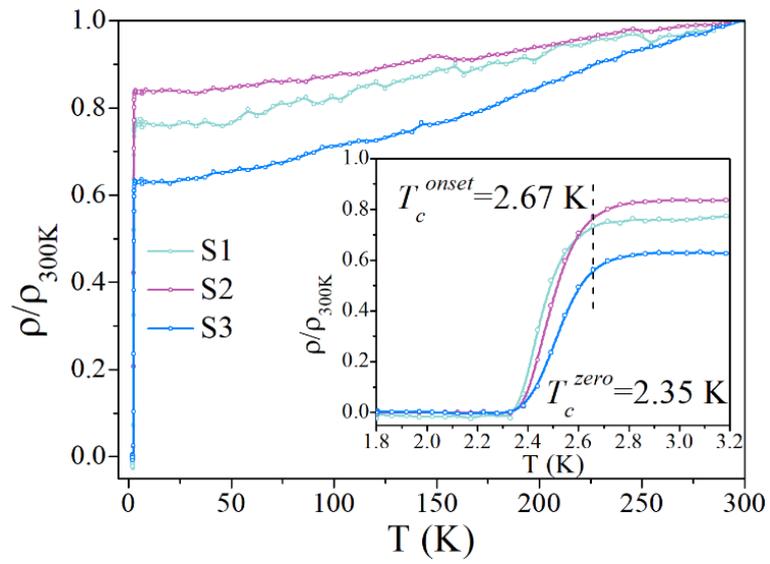

**Figure S3.** Temperature dependence of resistivity of three $(Ti_{0.2}Zr_{0.2}Hf_{0.2}Nb_{0.2}Ta_{0.2})C$ samples, where S1, S2, and S3 present sample 1, sample 2, and sample 3 respectively. The inset shows the detailed resistive transition. SC is observed with onset at ~2.67 K and zero resistance at ~ 2.35 K.

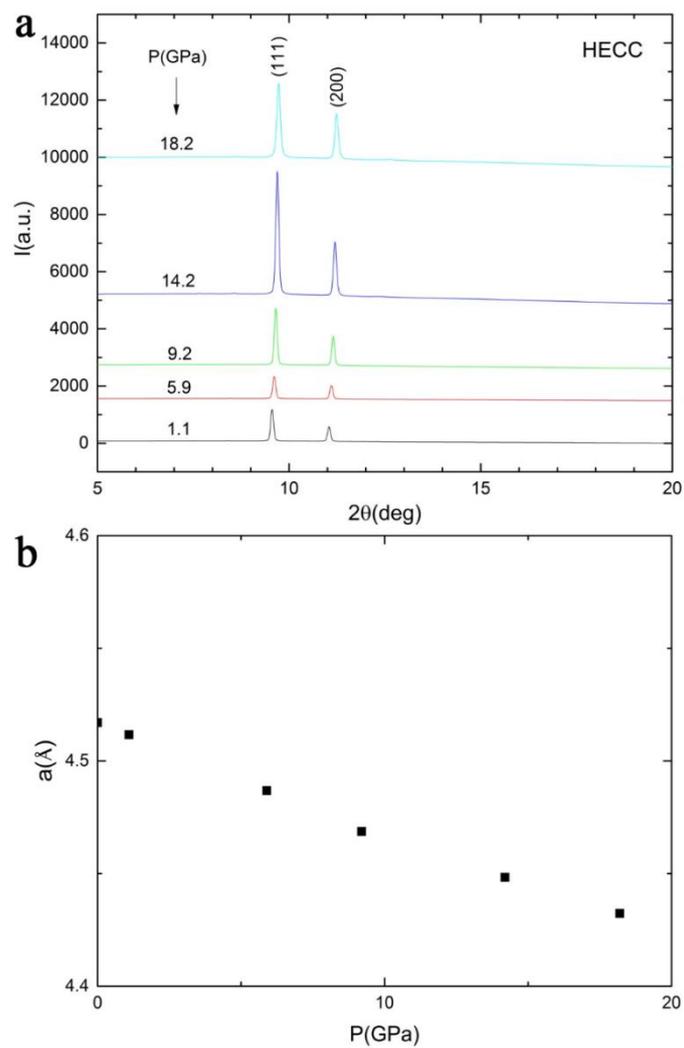

**Figure S4.** High-pressure (HP) XRD measurements on $(Ti_{0.2}Zr_{0.2}Nb_{0.2}Hf_{0.2}Ta_{0.2})C$. a) pressure-dependence of XRD patterns for $(Ti_{0.2}Zr_{0.2}Nb_{0.2}Hf_{0.2}Ta_{0.2})C$; b) the relationship between *a* parameter versus pressure.

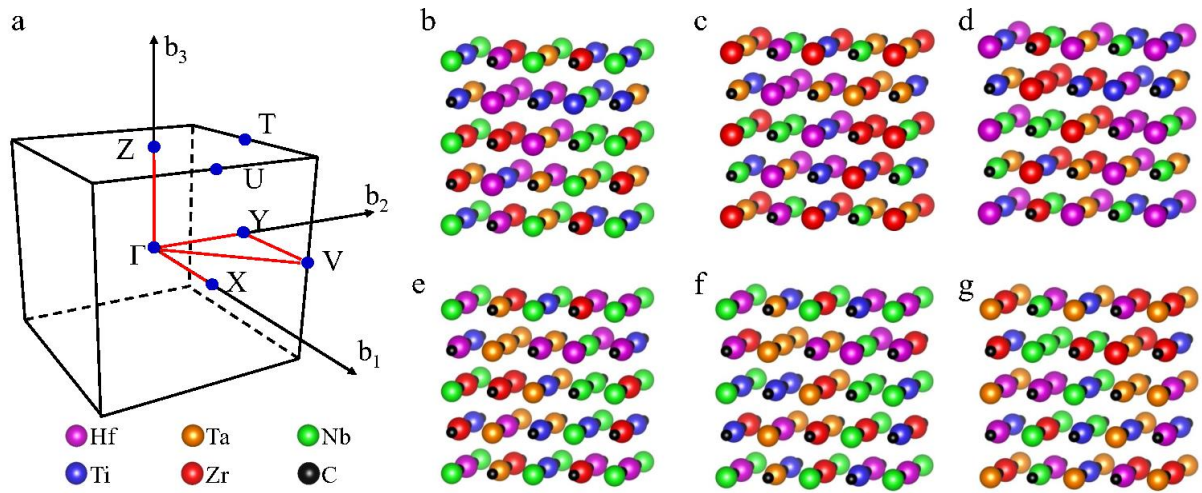

**Figure S5.** (a) The first Brillouin zone of $(Ti_{0.2}Zr_{0.2}Nb_{0.2}Hf_{0.2}Ta_{0.2})C$. The coordinates of Γ, X, Y, Z, V, U, T are Γ(0, 0, 0), X(0.5$b_1$, 0, 0), Y(0, 0.5$b_2$, 0), Z(0, 0, 0.5$b_3$), V(0.5$b_1$, 0.5$b_2$, 0), U(0.5$b_1$, 0, 0.5$b_3$), T(0, 0.5$b_2$, 0.5$b_3$). Here, $b_1$, $b_2$, and $b_3$ represent the basic axes in reciprocal space. (b)-(g) Crystal structures of the six representative SQSs for Figure 5 (a)-(f).

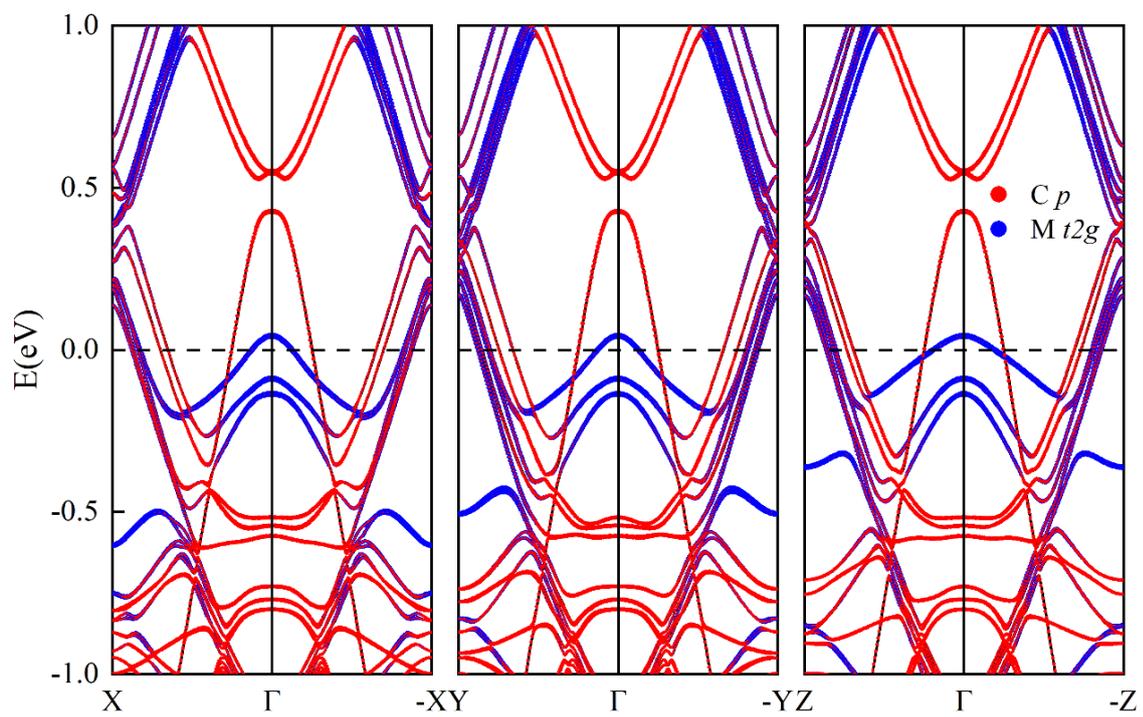

**Figure S6.** Six Dirac points in the $(Ti_{0.2}Zr_{0.2}Hf_{0.2}Nb_{0.2}Ta_{0.2})C$ with the crystal structure as shown in Figure S5 (b) along X-Γ--X, Y-Γ--Y, Z-Γ--Z.

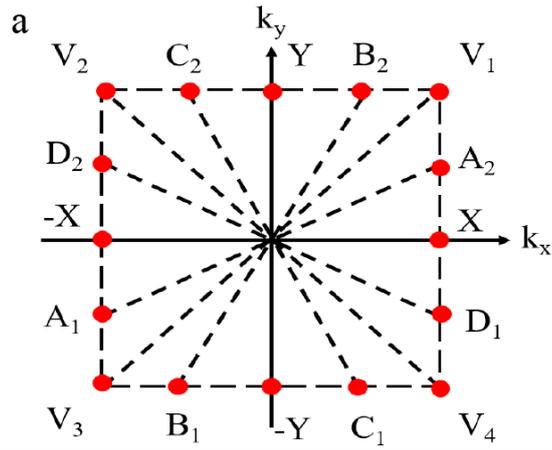
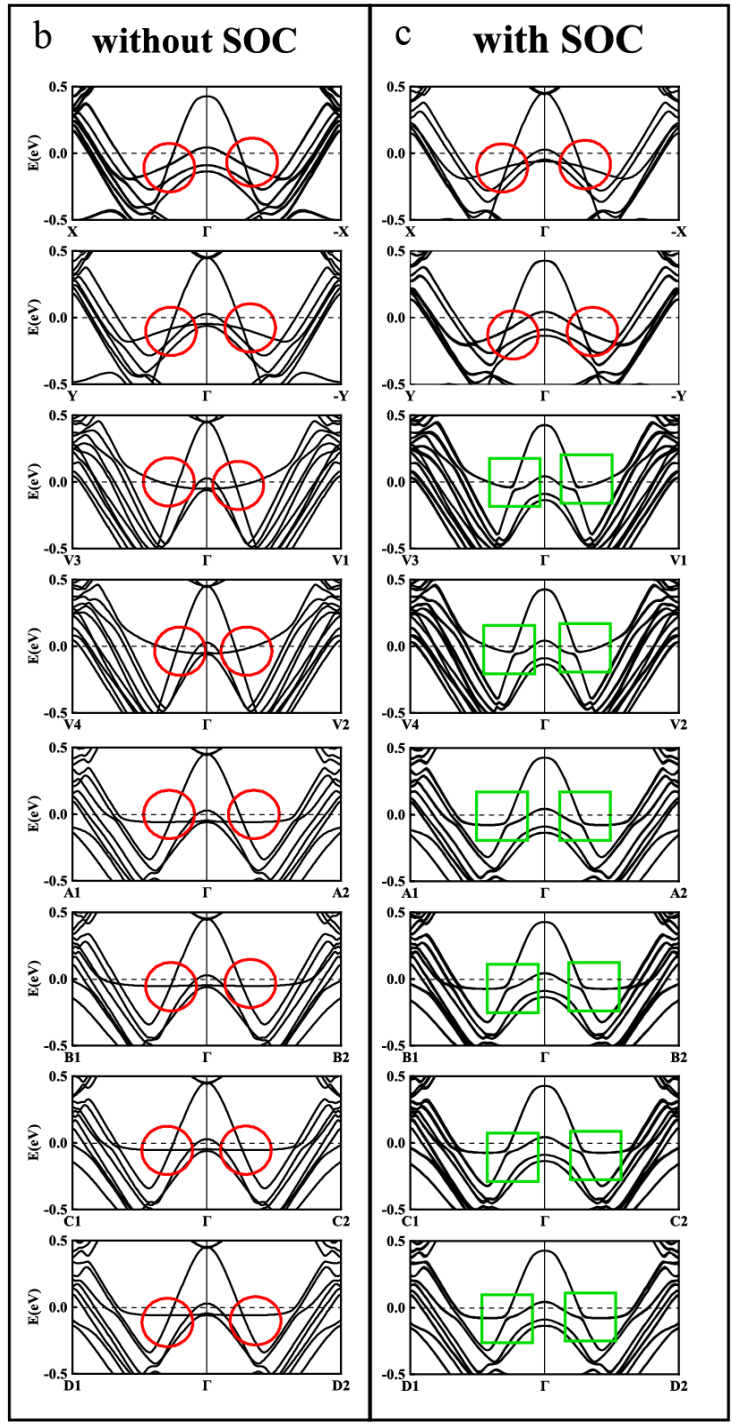

**Figure S7.** (a)The two-dimensional (2D) Brillouin zone of $(Ti_{0.2}Zr_{0.2}Nb_{0.2}Hf_{0.2}Ta_{0.2})C$ at $k_z$=0 plane. Band structures of $(Ti_{0.2}Zr_{0.2}Nb_{0.2}Hf_{0.2}Ta_{0.2})C$ (b) without SOC and (c) with SOC. Red circles and green squares represent band crossings and gapped bands, respectively. The coordinates of A$_1$, A$_2$, B$_1$, B$_2$, C$_1$, C$_2$, D$_1$, D$_2$ are A$_1$ (-0.5b$_1$, -0.25b$_2$, 0), A$_2$ (0.5b$_1$, 0.25b$_2$, 0), B$_1$ (-0.25b$_1$, -0.5b$_2$, 0), B$_2$ (0.25b$_1$, 0.5b$_2$, 0), C$_1$ (0.25b$_1$, -0.5b$_2$, 0), C$_2$ (-0.25b$_1$, 0.5b$_2$, 0), D$_1$ (0.5b$_1$, -0.25b$_2$, 0), D$_2$ (-0.5b$_1$, 0.25b$_2$, 0). Here, b$_1$, b$_2$, and b$_3$ represent the basic axes in reciprocal space.

From Figure S6b, we can obtain that $(Ti_{0.2}Zr_{0.2}Nb_{0.2}Hf_{0.2}Ta_{0.2})C$ has a nodal loop at the $k_z$=0 plane when SOC is not considered. However, when SOC is considered, this nodal loop will be gapped and four Dirac points appear at ($k_{x1}$, 0, 0), ($-k_{x2}$, 0, 0), (0, $k_{y1}$, 0) and (0, $-k_{y2}$, 0) (see Figure S6c). By combining Figures S6, S7, and S8, it comes to conclude that $(Ti_{0.2}Zr_{0.2}Nb_{0.2}Hf_{0.2}Ta_{0.2})C$ processes three nodal loops. These three nodal loops are located at $k_z$ = 0, $k_x$ = 0, and $k_y$ = 0 planes in the reciprocal space. When SOC is taken into account, these nodal loops are gaped and six Dirac points appear at ($k_{x1}$, 0, 0), ($-k_{x2}$, 0, 0), (0, $k_{y1}$, 0), (0, $-k_{y2}$, 0), (0, 0, $k_{z3}$) and (0, 0, $-k_{z3}$) in reciprocal space.

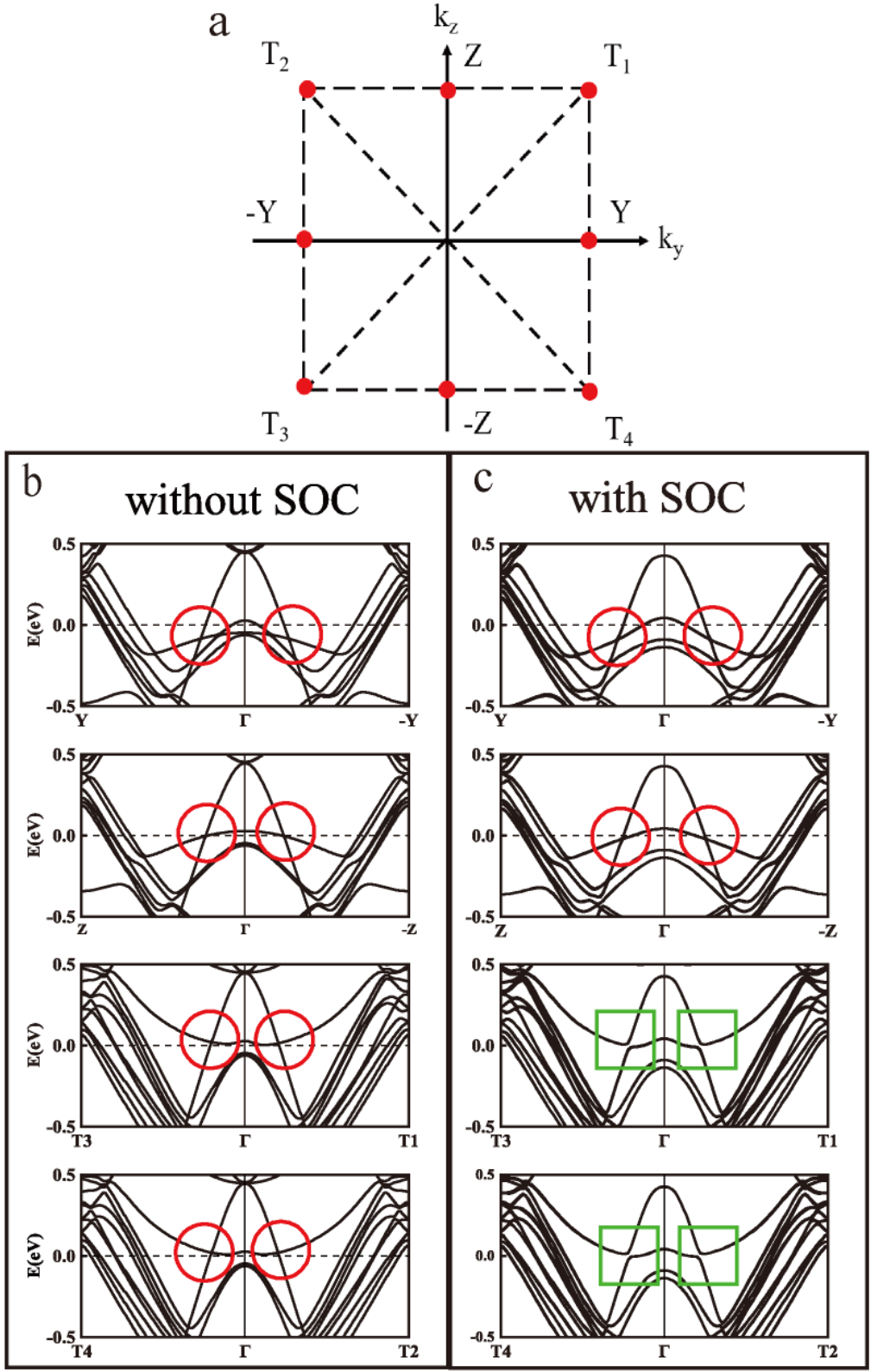

**Figure S8.** (a)The two-dimensional (2D) Brillouin zone of $(Ti_{0.2}Zr_{0.2}Nb_{0.2}Hf_{0.2}Ta_{0.2})C$ at $k_x=0$ plane. Band structures of $(Ti_{0.2}Zr_{0.2}Nb_{0.2}Hf_{0.2}Ta_{0.2})C$ (b) without SOC and (c) with SOC. Red circles and green squares represent band crossings and gapped bands, respectively. The coordinates of Y, Z, $T_1$, $T_2$, $T_3$, $T_4$, are $\Gamma(0, 0, 0)$, $Y(0, 0.5b_2, 0)$, $Z(0, 0, 0.5b_3)$, $T_1(0, 0.5b_2, 0.5b_3)$, $T_2(0, -0.5b_2, 0.5b_3)$, $T_3(0, -0.5b_2, -0.5b_3)$, $T_4(0, 0.5b_2, -0.5b_3)$. Here, $b_1$, $b_2$, and $b_3$ represent the basic axes in reciprocal space.

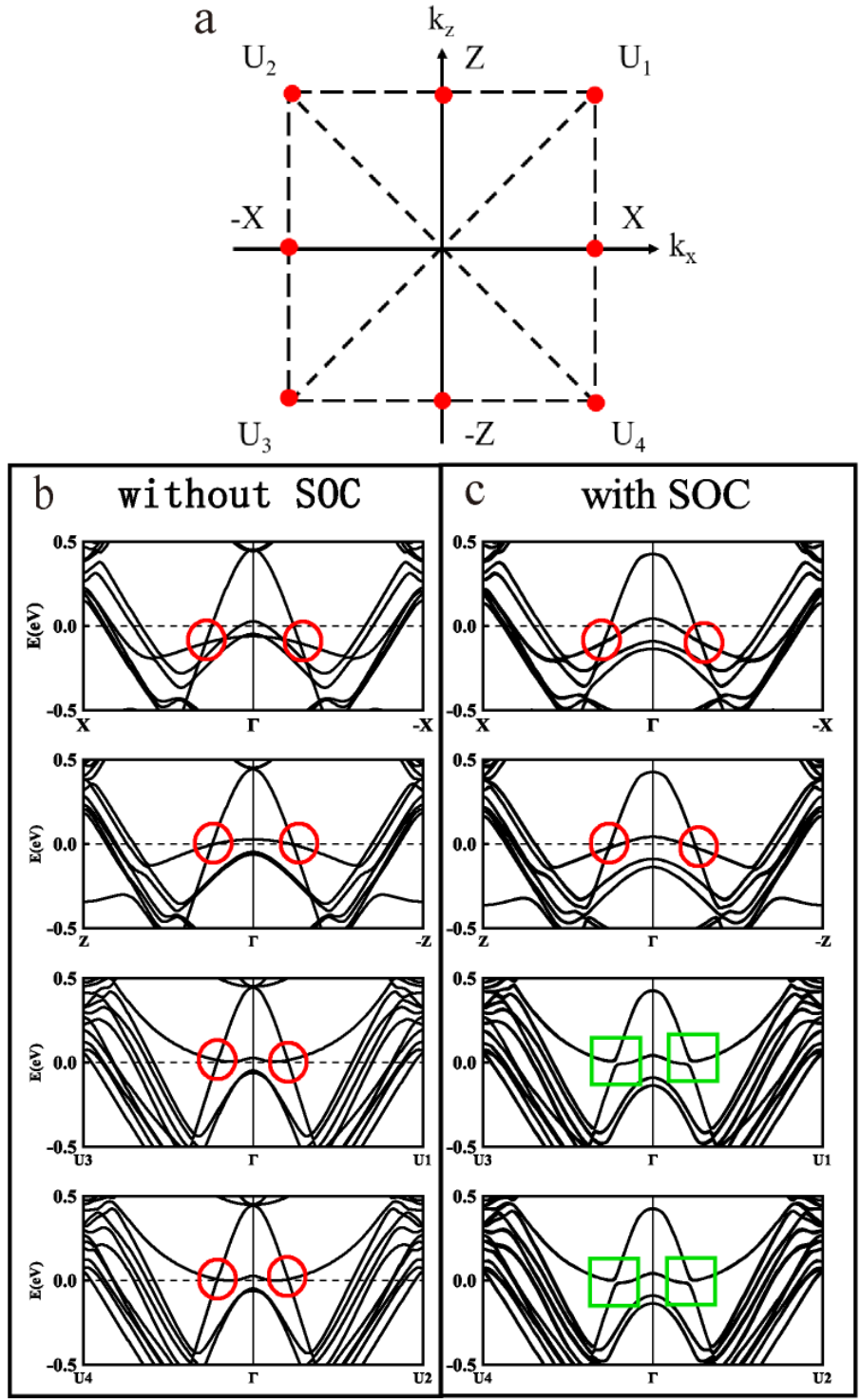

**Figure S9.** (a)The two-dimensional (2D) Brillouin zone of $(Ti_{0.2}Zr_{0.2}Nb_{0.2}Hf_{0.2}Ta_{0.2})C$ at $k_y$=0 plane. Band structures of $(Ti_{0.2}Zr_{0.2}Nb_{0.2}Hf_{0.2}Ta_{0.2})C$ (b) without SOC and (c) with SOC. Red circles and green squares represent band crossings and gapped bands, respectively. The coordinates of X, Z, $U_1$, $U_2$, $U_3$, $U_4$, are X(0.5$b_1$, 0, 0), Z(0, 0, 0.5$b_3$), $U_1$(0.5$b_1$, 0, 0.5$b_3$), $U_2$(-0.5$b_1$, 0, 0.5$b_3$), $U_3$(-0.5$b_1$, 0, -0.5$b_3$), $U_4$(0.5$b_1$, 0, -0.5$b_3$). Here, $b_1$, $b_2$, and $b_3$ represent the basic axes in reciprocal space.

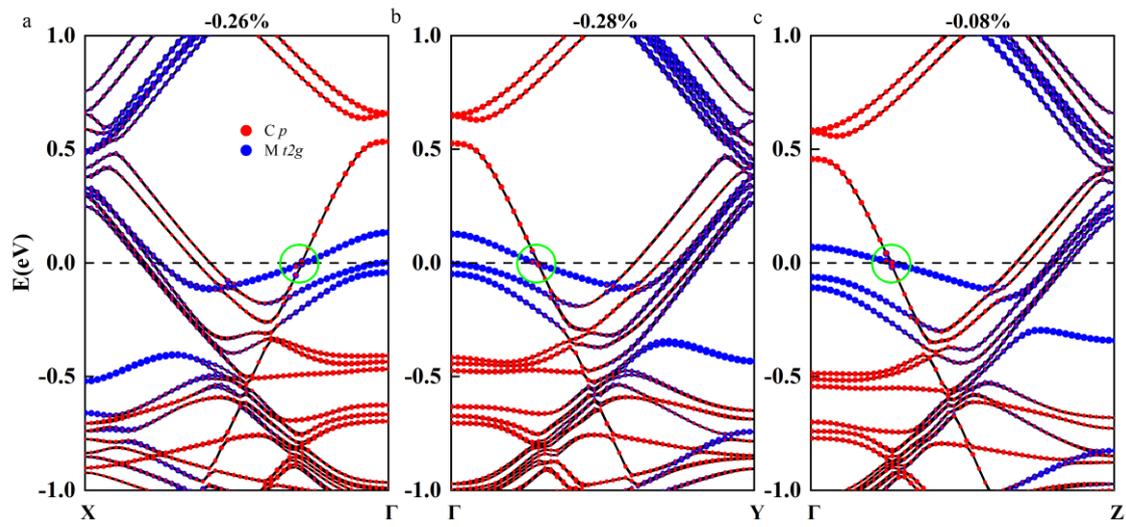

**Figure S10.** The orbital-resolved electronic band structures of $(Ti_{0.2}Zr_{0.2}Nb_{0.2}Hf_{0.2}Ta_{0.2})C$ under strain (a) -0.26%, (b) -0.28%, (c) -0.08%.

**Table S1**. The superconductivity parameters of $(Ti_{0.2}Zr_{0.2}Nb_{0.2}Hf_{0.2}Ta_{0.2})C$, NbC [25], and TaC [25].

| Property | Unit | NbC | TaC | $(Ti_{0.2}Zr_{0.2}Nb_{0.2}Hf_{0.2}Ta_{0.2})C$ |
|---|---|---|---|---|
| $T_c^\chi$ | K | 11.5 | 10.3 | 2.35 |
| $T_c^\rho$ | K | 10.6 | 10.2 | 2.51 |
| $\mu_0 H_{c1}$ | mT | 10.3 | 29.3 | 26.1 |
| $\mu_0 H_{c2}$ | T | 1.93 | 0.65 | 0.51 |
| $\mu_0 H_{c2}/T_c^\chi$ | T/K | 0.168 | 0.063 | 0.217 |
| $\xi_{GL}(0)$ | Å | 131.1 | 225.1 | 261.84 |
| $\gamma$ | mJ·mol$^{-1}$·K$^{-2}$ | 3.0(4) | 3.1(8) | 1.85 |
| $\Theta_D$ | K | 790 | 590 | 724 |
| $\lambda_{ep}$ |  | 0.60 | 0.65 | 0.45 |
| $N(E_F)$ | states/eV f.u. | 0.8 | 0.8 | 0.54 |